\documentclass{iopart}

\usepackage{bm}
\usepackage{amsfonts}
\begin{document}
\newcommand{\drawsquare}[2]{\hbox{%
\rule{#2pt}{#1pt}\hskip-#2pt
\rule{#1pt}{#2pt}\hskip-#1pt
\rule[#1pt]{#1pt}{#2pt}}\rule[#1pt]{#2pt}{#2pt}\hskip-#2pt
\rule{#2pt}{#1pt}}

\newcommand{\Yfund}{\raisebox{-.5pt}{\drawsquare{6.5}{0.4}}}
\newcommand{\Yasymm}{\raisebox{-3.5pt}{\drawsquare{6.5}{0.4}}\hskip-6.9pt%
                     \raisebox{3pt}{\drawsquare{6.5}{0.4}}%
                    }
\newcommand{\Ysymm}{\Yfund\hskip-0.4pt%
                    \Yfund}
\def\symm{\Ysymm}
\def\bsymm{\overline{\Ysymm}}
\def\ls{\mathrel{\lower4pt\vbox{\lineskip=0pt\baselineskip=0pt
           \hbox{$<$}\hbox{$\sim$}}}}
\def\gs{\mathrel{\lower4pt\vbox{\lineskip=0pt\baselineskip=0pt
           \hbox{$>$}\hbox{$\sim$}}}}
\def\drawbox#1#2{\hrule height#2pt
        \hbox{\vrule width#2pt height#1pt \kern#1pt
              \vrule width#2pt}
              \hrule height#2pt}

\def\Fund#1#2{\vcenter{\vbox{\drawbox{#1}{#2}}}}
\def\Asym#1#2{\vcenter{\vbox{\drawbox{#1}{#2}
              \kern-#2pt       
              \drawbox{#1}{#2}}}}
\def\sym#1#2{\vcenter{\hbox{ \drawbox{#1}{#2} \drawbox{#1}{#2}    }}}
\def\fund{\Fund{6.4}{0.3}}
\def\asymm{\Asym{6.4}{0.3}}
\def\bfund{\overline{\fund}}
\def\basymm{\overline{\asymm}}


\newcommand{\beq}{\begin{equation}}
\newcommand{\eeq}{\end{equation}}
\def\ls{\mathrel{\lower4pt\vbox{\lineskip=0pt\baselineskip=0pt
           \hbox{$<$}\hbox{$\sim$}}}}
\def\gs{\mathrel{\lower4pt\vbox{\lineskip=0pt\baselineskip=0pt
\def\lsim{\mathrel{\lower4pt\vbox{\lineskip=0pt\baselineskip=0pt
           \hbox{$<$}\hbox{$\sim$}}}}
\def\gsim{\mathrel{\lower4pt\vbox{\lineskip=0pt\baselineskip=0pt
           \hbox{$>$}\hbox{$\sim$}}}}           \hbox{$>$}\hbox{$\sim$}}}}

\title{Identifying the curvaton within MSSM }

\author{Rouzbeh Allahverdi$^{1,2}$, Kari Enqvist$^{3}$, Asko Jokinen$^{4}$ and Anupam Mazumdar$^{4}$}
\address{$^{1}$~Perimeter Institute for Theoretical Physics, Waterloo, ON,
N2L 2Y5, Canada. \\
$^{2}$~Department of Physics and Astronomy, McMaster University, Hamilton, 
ON, L8S 4M1, Canada. \\ 
$^{3}$~Department of Physical Sciences, University of Helsinki,
and Helsinki Institute of Physics,
P.O. Box 64, FIN-00014 University of Helsinki, Finland \\
$^{4}$~NORDITA, Blegdamsvej-17, Copenhagen-2100, Denmark.}
\eads{\mailto{rallahverdi@perimeterinstitute.ca}, \mailto{kari.enqvist@helsinki.fi}, \mailto{ajokinen@nordita.dk}, \mailto{anupamm@nordita.dk}}

\begin{abstract}
We consider inflaton couplings to MSSM flat directions and the
thermalization of the inflaton decay products, taking into account
gauge symmetry breaking due to flat direction condensates.  We then
search for a suitable curvaton candidate among the flat directions,
requiring an early thermally induced start for the flat direction
oscillations to facilitate the necessary curvaton energy density
dominance. We demonstrate that the supersymmetry breaking $A$-term is
crucial for achieving a successful curvaton scenario. Among the many
possible candidates, we identify the ${\bf u_1dd}$ flat direction as a
viable MSSM curvaton.
\end{abstract}
\maketitle

\section{Introduction}

Minimal Supersymmetric Standard Model (MSSM) provides nearly $300$
gauge in\-va\-riant flat directions, whose potentials are vanishing in a
perfect supersymmetric (SUSY) limit (for a review, see
\cite{Enqvist-REV}).  However, in the early Universe SUSY is broken
and the flatness of the potentials is spoiled, but the directions
still remain flat as compared to the curvature scalar of the Universe.
Soft SUSY breaking terms induce $m_0\sim {\cal O}(\rm TeV)$ masses to
the flat direction.  Flatness is also lifted by non-renormalizable
superpotential terms, the form of which is dictated by the gauge
properties of the flat direction.  When the Hubble expansion rate
becomes equal to the low energy SUSY scale, all the flat directions
are trapped in their local minima, after which they start to oscillate
and ultimately decay.  The flat directions may a play a crucial role
in cosmological issues ranging from baryogenesis to dark generation
matter. Recently it has also been pointed out that they are also
important for understanding the full thermalization history of the
Universe after inflation~\cite{Averdi1,Averdi2}.

An MSSM flat direction may also account for generating adiabatic
density perturbations~\cite{Enqvist1,Enqvist2,Enqvist3,Postma} through
the curvaton mechanism~\cite{Sloth,Lyth-Wands,Moroi}.  Since during
inflation the Hubble expansion rate $H_I \gg m_0$ it does not cost
anything in energy, quantum fluctuations are free to accumulate (in a
coherent state) along a flat direction and form a condensate with a
large VEV, $\varphi_0$.  Because inflation smoothes out all the
gradients, only the homogeneous condensate mode survives. However, the
zero point fluctuations of the condensate impart a small, and in
inflationary models a calculable, spectrum of perturbations on the
condensate.

After inflation, $H \propto t^{-1}$ and the flat direction stays at a
relatively large VEV due to large Hubble friction term until $H \simeq
m_0$, whence the flat direction condensate starts oscillating around
the origin with an initial amplitude $\sim \varphi_0$. From then on
$\vert \varphi \vert$ is redshifted by the Hubble expansion $\propto
H$ for matter dominated and $\sim H^{3/4}$ for radiation dominated
Universe. The energy of the oscillating flat direction may eventually
start to dominate over the inflaton decay products.  When the flat
direction decays, its isocurvature perturbations will be converted to
the usual adiabatic perturbations of the decay products, which thus
should ultimately contain also Standard Model (SM) degrees of freedom.
However, such an evolution is not automatic but is subject to various
constraints \cite{Enqvist1,Enqvist2}, with the outcome that in general
it is very difficult to have a MSSM flat direction curvaton that could
give rise to the desired spectrum of density perturbations.

The key ingredient in these considerations is the process of
thermalization of the ambient plasma.  The state of the plasma before
complete thermalization depends on the nature of flat direction which
is developing a large VEV. For instance, MSSM Higgses developing a VEV
give masses to the gauge bosons of $SU(2)_{\rm W}\times U(1)_{\rm Y}$,
but gluons remain massless. The decay products can then reach partial
thermalization through the gluons. Therefore one needs to study
various possibilities case by case. It is important to know the
details of the inflaton couplings to the MSSM degrees as well, since
not all the couplings are renormalizable.

Thermal effects are important for the MSSM curvaton
mechanism~\cite{Thermal1,Thermal2} because large thermal corrections
can evaporate or dissociate the MSSM curvaton candidate before it has
a chance to dominate the energy density.  These effects obviously
depend on the thermalization rate.  Recently it has been pointed out
that the VEVs of the gauge invariant flat directions actually slow
down the thermalization rate of the inflaton decay
products~\cite{Averdi1,Averdi2}. This is due to the fact that, in
general, the flat direction VEV gives masses to gauge bosons and
gauginos which decrease the rates governed by $2\leftrightarrow 2$ and
$2\rightarrow 3$ scatterings mediated via gauge bosons and
gauginos. After the inflaton decay~\footnote{Within SUSY the inflaton
predominantly decays perturbatively. Non-perturbative decay of the
inflaton to MS(SM) or through selfcoupling~\cite{EKM} via parametric
resonance is kinematically blocked because the inflaton decay products
can couple (in a gauge invariant fashion) to MSSM flat
directions. Large VEV of the flat direction induces large masses to
the inflaton decay products which blocks preheating
completely~\cite{AM}.}  the initial plasma will be in a quasi-thermal
phase for a long period during which kinetic equilibrium is
established but not the full chemical equilibrium and the initial
plasma is far from thermal equilibrium.

Another consequence of thermal corrections is that if the flat
direction survives long enough, then thermally induced mass
corrections can trigger early oscillations, which may hasten the
dominance of the curvaton over the inflaton decayed products. These
subtleties has never been systematically considered in the context of
an MSSM curvaton. It is the purpose of the present paper to make use
of the results of ref.~\cite{Averdi1,Averdi2} and study these effects
properly.

In previous work the evolution of the MSSM curvaton has been
discussed~\cite{Enqvist1,Enqvist2} without due consideration of the
SUSY breaking $A$-term in the potential, which is induced by the
presence of the non-renormalizable terms in the superpotential.  Here
we take the $A$-terms into account and show that they greatly modify
the dynamical evolution and in particular the initial amplitude of the
curvaton oscillations. As a consequence, many of the obstacles for an
MSSM flat direction curvaton are now removed, and considering the
various suitable directions case by case we are able to pick out the
${\bf u_1dd}$ flat direction as the most promising curvaton candidate.

The paper is organized as follows.  In Section~\ref{Infd} we discuss
how the inflaton couples to the MSSM degrees of freedom, emphasizing
that there are certain combinations of slepton and squark superfields
that do not have renormalizable couplings to a gauge singlet
inflaton. In Section~\ref{Curvatonc} we disseminate the potential flat
direction curvaton candidates and choose those which give rise to
thermal effects that can trigger early curvaton oscillations.  Section
\ref{CPA} explains the importance of the $A$-term for MSSM curvaton
dynamics, while in Section \ref{TCF} we discuss the thermal
corrections to flat directions, which singles out the ${\bf udd}$
direction as the curvaton.  In Section~\ref{udd} we discuss the
dynamics of the ${\bf udd}$ curvaton and find its initial amplitude
and a lower bound on the renormalizable inflaton coupling. Our
conclusions are presented in Section~\ref{lopetus}.

\section{Inflaton decay}
\label{Infd}

In almost all $F$-term or $D$-term models of inflation the inflaton,
$\Phi$, is considered to be an absolute gauge singlet~\footnote{One
exceptional example is ref.~\cite{Asko-Assist}, where the gauge
invariant combinations of SO(N) flat directions are responsible for
driving inflation.}. Then the pertinent question is how the inflaton
couples to the matter.

Gauge symmetry implies that the inflaton must be coupled to a
gauge-invariant combination of the SM fields and their SUSY
partners. The field content of MSSM is governed by the following
superpotential:
\begin{equation}
\label{mssm}
W_{MSSM}=\lambda_u{\bf Q} {\bf H_u} {\bf u} + \lambda_d {\bf Q} {\bf H_d}
{\bf d} + \lambda_e {\bf L} {\bf H_d} {\bf e}~
+ \mu {\bf H_u} {\bf H_d}\,,
\end{equation}\
where ${\bf H_u}, {\bf H_d}, {\bf Q}, {\bf L}, {\bf u}, {\bf d}, {\bf
e}$ in Eq.~(\ref{mssm}) are chiral superfields representing the two
Higgs fields (and their Higgsino partners), LH (s)quark doublets, RH
up- and down-type (s)quarks, LH (s)lepton doublets and RH (s)leptons
respectively.  The dimensionless Yukawa couplings $\lambda_{u},
\lambda_{d}, \lambda_{e}$ are $3\times 3$ matrices in the flavor
space, and we have omitted the gauge and flavor indices. The last term
is the $\mu$ term, which is a supersymmetric version of the SM Higgs
boson mass.

There exist two gauge-invariant combinations of only two superfields:
\beq \label{two}
{\bf H_u} {\bf H_d} ~ , ~ {\bf H_u} {\bf L}.
\eeq
The combinations which include three superfields are:
\beq \label{three}
{\bf H_u} {\bf Q} {\bf u} , ~ {\bf H_d} {\bf Q} {\bf d} , ~ {\bf H_d}
{\bf L} {\bf e} ~, ~ {\bf Q} {\bf L}
{\bf d} , ~ {\bf u} {\bf d} {\bf d} , ~ {\bf L} {\bf L} {\bf e}.
\eeq
SUSY together with gauge symmetry requires that the inflaton
superfield ${\bf \Phi}$ is coupled to these combinations or to their
superpositions. The terms ${\bf \Phi} {\bf H_u} {\bf H_d}$ and ${\bf
\Phi} {\bf H_u} {\bf L}$ have dimension four, and hence are
renormalizable. On the other hand, the interaction terms that couple
the inflaton to the combinations in Eq.~(\ref{three}) have dimension
five and are non-renormalizable. They can arise after integrating out
the heavy degrees of freedom which are associated to the new physics
at high scales (for example GUT). Such terms are also generically
induced by gravity, notably in supergravity, in which case they will
be Planck mass suppressed~\footnote{It is possible that the inflaton
decays mainly to another singlet (for example, the right-handed
neutrino) superfield. Then, since the ultimate goal is to create
matter, this singlet must be coupled to MSSM fields.  Supersymmetry
and gauge symmetry again require couplings to gauge-invariant
combinations in Eqs.~(\ref{two}) and~(\ref{three}) along the lines
above.}.

The simplest case arises when the inflaton is coupled to matter via
superpotential terms of the form:
\begin{eqnarray} \label{inf}
h {\bf \Phi} {\bf H_u} {\bf H_d}  \, ,
h {\bf \Phi} {\bf H_u} {\bf \Psi} \, ,
\end{eqnarray}
where we have defined
\begin{equation} \label{def}
{\bf \Psi} = \frac{1}{h} \sum_{i}{h_i {\bf L}_i} ~ , ~ ~
h = \left[\sum_{i}{\left(h_i\right)^2}\right]^{1/2}~.
\end{equation}
Here $1 \leq i \leq 3$ is a flavor index and $h$ can be as large as
${\cal O}(1)$.

It is important to realize that one has to go from the $\left\{{\bf
L}_1,{\bf L}_2,{\bf L}_3 \right\}$ basis, in which $\lambda_e$ in
Eq.~(\ref{mssm}) are diagonalized, to the $\left\{{\bf \Psi},{\bf
L^{\prime}}_1,{\bf L^{\prime}}_2\right\}$, where ${\bf L^{\prime}}_1$
and ${\bf L^{\prime}}_2$ are orthogonal to ${\bf \Psi}$, and hence
have no renormalizable couplings to the inflaton; only ${\bf \Psi}$
couples to the inflaton.  (for a discussion on multi-flat directions,
see ~\cite{Asko-multi}).  This is just a unitary transformation.

Besides the SM gauge group the MSSM Lagrangian is also invariant under
R-parity ${(-1)}^{3B + L + 2S}$.  Preserving R-parity (at least) at
the renormalizable level further constrains inflaton couplings to
matter. Note that ${\bf H}_u {\bf H}_d$ is assigned $+1$ under
R-parity, while ${\bf H}_u {\bf \Psi}$ has the opposite assignment
$-1$. Therefore only one of the couplings in Eq.~(\ref{inf}) preserves
R-parity: ${\bf \Phi} {\bf H}_u {\bf H}_d$ if the inflaton superfield
carries no lepton or baryon number, which is generically the case. In
contrast, the ${\bf \Phi} {\bf H}_u {\bf \Psi}$ term could be relevant
for e.g. models where the RH sneutrino plays the role of the
inflaton~\cite{sninfl}.

In the context of a curvaton scenario the fact that the inflaton can
have renormalizable couplings to (some of) the MSSM fields is an
auspicious sign.  The interactions in Eq.~(\ref{inf}) result in an
inflaton decay rate:
\begin{equation}
\label{Gammad}
\Gamma_{\rm d} = \left(\frac{h^2}{4 \pi}\right) m_{\phi} .
\end{equation}
On the other hand, non-renormalizable interactions from
Eq.~(\ref{three}) result in a much smaller decay rate $\sim
m^3_{\phi}/M^2_{\rm P}$. Such a late decay would imply a longer period
of inflaton domination during which the ratio of the curvaton energy
density to the total energy density of the Universe does not
grow. Moreover, a small decay rate would result in a low reheat
temperature, and hence would diminish the thermal effects necessary to
trigger early oscillations of the curvaton field, as will be discussed
below.

\section{Curvaton candidates}
\label{Curvatonc}

Having discussed the inflaton couplings to matter, let us now attempt
to identify the curvaton candidates within MSSM. An important point is
that generating density perturbations of the correct size from
curvaton decay requires that during inflation its VEV $\varphi_0\sim
10^5 H_{\rm I}$. Here $H_{\rm I}$ is the Hubble expansion rate during
the inflationary epoch.

A curvaton is a light modulus, lighter than the Hubble expansion rate
during inflation. The total potential during inflation is given by
\begin{equation}
V_{total}=V(I)+V(\varphi)\,
\end{equation}
where $V(I)$ is due to inflation and $V(\varphi)$ is due to the
curvaton. The curvaton carries isocurvature perturbations which
sources the curvature perturbations. In order not to have any residual
isocurvature perturbations left over the curvaton must decay
dominantly into the SM degrees of freedom.  The interesting quantity
to study is the ratio of the perturbation and the background field
value of the curvaton, since this is related to the curvature
perturbation ~\cite{Sloth,Lyth-Wands}, which can be constrained from
the amplitude of the CMB anisotropy. If the perturbations in the
curvaton do not damp then the final curvature perturbation will be
given by
\begin{equation}
\label{cobe}
\delta = \frac{H_{inf}}{\varphi_{inf}}\sim 10^{-5}\,,
\end{equation}
where $10^{-5}$ arises from the COBE normalization. In this we will
present a scenario where the perturbations are not damped away and the
curvaton dominates while decaying.

There are three important points to note:

\begin{itemize}

\item{\it The flat direction that plays the role of the curvaton
cannot include those fields which have a renormalizable coupling to
the inflaton. The reason is that such fields generically acquire a
mass $h \phi \gg H_{inf}$ through their couplings to the inflaton, and
hence cannot develop the required VEV. Therefore Eq.~(\ref{inf})
implies that a curvaton candidate cannot include $H_u$, $H_d$ and
$\Psi$. As pointed out earlier, such a late decay would be harmful to the
curvaton scenario.}

\item{\it A curvaton candidate should not induce a mass $\geq
m_{\phi}/2$ for the inflaton decay products, otherwise the two-body
inflaton decay will be kinematically blocked.  The decay will be
delayed until the relevant flat direction has started its oscillation
and its VEV has been redshifted to sufficiently small
values~\cite{AM}.}

\item{\it The flat direction which plays the role of the curvaton
should not break all of the SM gauge symmetry. The reason is that in
this case all gauge integrations will get decoupled and full (i.e.
both kinetic and chemical) thermal equilibrium will be established
much later than the inflaton decay~\cite{Averdi2}~\footnote{If only
part of the SM group is broken, gauge interactions of the unbroken
part will bring fields which carry those gauge quantum numbers, and
subsequently other fields, into full equilibrium.}. However thermal
effects which are necessary to trigger early oscillations of the flat
direction~\cite{Thermal1,Thermal2} crucially depend on the presence of
a thermal bath consisting of some MSSM degrees of freedom in full
equilibrium when the inflaton decay is completed.}

\end{itemize}

Note that a flat direction, which is a combination of the squark and
slepton fields, has couplings to other MSSM fields through which it
induces a large mass for them. D-terms lead to flat direction
couplings of gauge strength to the gauge fields and gauginos
associated with the gauge (sub)group which is spontaneously broken by
the flat direction VEV, as well as to the orthogonal directions and
their supersymmetric partners.  F-terms result in flat direction
couplings of Yukawa strength to the scalars that are not included in
the monomial representing the flat direction, as well as to their
fermionic partners.

To elucidate these points, let us consider the particularly simple
example of the $H_u H_d$ flat direction\footnote{Note that, as we
mentioned, this flat direction cannot obtain a large VEV. We only
consider this example to demonstrate how the flat direction VEV
induces mass to other MSSM fields.}. In this case the flat direction
and the orthogonal directions are defined as $\left(H_{u,1} + H_{d,2}
\right)/\sqrt{2}$, and $\left(H_{u,1} - H_{d,2} \right)/\sqrt{2}$,
$H_{u,2}$ and $H_{d,1}$, respectively, where $1,2$ are the two
components of $H_{u}$ and $H_d$.  The flat direction breaks $SU(2)_W
\times U(1)_Y$ down to $U(1)_{\rm em}$. Then the gauge fields and
gauginos associated with the broken subgroup, as well as
$\left(H_{u,1} - H_{d,2} \right)/\sqrt{2},~H_{u,2},~H_{d,1}$ and their
fermionic partner, acquire a mass $\simeq g \langle\varphi\rangle$,
where $g$ is a gauge coupling. The ${\bf Q}_i$ multiplets, which are
coupled to both $H_u$ and $H_d$, acquire a mass
$\sqrt{(\lambda^2_{u,i} + \lambda^2_{d,i})} \langle \varphi\rangle$
while ${\bf u}_i$, ${\bf d}_i$ and ${\bf e}_i$ multiplets obtain
respectively masses $\lambda_{u,i}\langle \varphi\rangle$,
$~\lambda_{d,i} \langle\varphi\rangle$, and $\lambda_{e,i}\langle
\varphi\rangle$.

Let us denote the flat direction superfield by $\varphi$. Then we have
\beq \label{flatcoupling}
W_{\rm MSSM} \supset \lambda_1 {\bf H}_u \varphi {\bf \Sigma}_1 + \lambda_2
{\bf H}_d \varphi {\bf \Sigma}_2 + \lambda_3 {\bf \Psi} \varphi {\bf \Sigma}_3,
\eeq
where ${\bf \Sigma}_{1,2,3}$ are MSSM superfields.  In general the
relationship $m_{\phi} \leq H_{inf}$ holds~\footnote{This is strictly
true in models of chaotic inflation and (supersymmetric) hybrid
inflation. It is possible to have $m_{\phi} \gg H_{inf}$ in models of
new inflation, but this is a rather contrived situation.}.  For the
curvaton mechanism to work, we require $\varphi_{inf} \sim 10^5
H_{inf}$. Note that the VEV of the flat direction induces VEV
dependent SUSY preserving masses to the MSSM particles.  Therefore,
for the inflaton decay to be kinematically allowed, one needs
\beq \label{yukawabound1}
\lambda_{1},\lambda_2 \leq 10^{-5},
\eeq
if the ${\bf \Phi} {\bf H}_u {\bf H}_d$ coupling is allowed by R-parity, and
\beq \label{yukawabound2}
\lambda_1,\lambda_3 \leq 10^{-5},
\eeq
if the ${\bf \Phi} {\bf H}_u {\bf \Psi}$ coupling is allowed.

The above conditions considerably restrict the curvaton candidates
within the MSSM as only the first generations of (s)quarks and
(s)leptons which have a Yukawa coupling $\ls 10^{-5}$.

Let us now identify the flat directions which satisfy all of the above
mentioned requirements. An important point is that the acceptable flat
directions should include {\it only} one ${\bf Q}$ or one ${\bf
u}$. The reason is that $D$- and $F$-flatness of directions which
involve two or more ${\bf Q}$ and/or ${\bf u}$ requires them to be of
different flavors (for details, see \cite{gkm}). This implies the
presence of up-type squarks from the second and/or third generation
which, according to Eq.~(\ref{flatcoupling}), will lead to $\lambda_1
\geq 10^{-3}$. This violates the condition for two-body inflaton decay
via either of the ${\bf \Phi} {\bf H}_u {\bf H}_d$ or ${\bf \Phi} {\bf
H}_u {\bf \Psi}$ terms, given in
Eqs.~(\ref{yukawabound1},\ref{yukawabound2}).  A large number of MSSM
flat directions will therefore be excluded by this consideration. The
only flat directions with $\lambda_1 \leq 10^{-5}$ are as follows:
\vskip5pt
\noindent
\begin{itemize}

\item{$\underline{{\bf u} {\bf d} {\bf d}}$: This monomial represents
a subspace of complex dimension $6$~\cite{gkm}. $D$-flatness requires
that the two ${\bf d}$ are from different generations (hence at least
one of them will be from the second or third generation). This implies
that $\lambda_2 \geq 10^{-3}$, see Eq.~(\ref{flatcoupling}), for all
flat directions classified by ${\bf u} {\bf d} {\bf d}$. As a
consequence the two-body inflaton decay via the ${\bf \Phi} {\bf H}_u
{\bf H}_d$ term will be kinematically forbidden. Note however that
these flat directions are not coupled to ${\bf L}_{1,2,3}$. Therefore
the inflaton decay via the ${\bf \Phi} {\bf H}_u {\bf \Psi}$ term can
proceed for the ${\bf u}_1 {\bf d} {\bf d}$ directions, with $u_1$
being the RH up squark. We also note that these directions leave the
$SU(2)_{\rm W}$ unbroken, so that the $SU(2)_{\rm W}$ degrees of
freedom can completely thermalize.}
\vskip5pt
\item{$\underline{{\bf Q} {\bf L} {\bf d}}$: This monomial
represents a subspace of complex dimension $19$~\cite{gkm}. $F$-flatness
requires that ${\bf Q}$ and ${\bf d}$ belong to different
generations. Then, since ${\bf Q}$ and ${\bf d}$ are both coupled to
${\bf H}_d$, Eq.~(\ref{flatcoupling}) implies that $\lambda_2 \geq
10^{-3}$.  Thus two-body inflaton decay via the ${\bf \Phi} {\bf H}_u
{\bf H}_d$ term will be kinematically forbidden. On the other hand it
can proceed via the ${\bf \Phi} {\bf H}_u {\bf \Psi}$ term for ${\bf
Q}_1 {\bf L^{\prime}} {\bf d}$ directions, where $Q_1$ is the doublet
containing LH up and down squarks.  Note that here we have to rotate
to the $\left\{{\bf \Psi},{\bf L^{\prime}}_1, {\bf
l^{\prime}}_2\right\}$ basis where only $L^{\prime}_1$ and
$L^{\prime}_2$ can acquire a large VEV during inflation. We also note
that these directions completely break the $SU(2)_{\rm W} \times
U(1)_{\rm Y}$, but leave a $SU(2)$ subgroup of the $SU(3)_{\rm C}$
unbroken. Therefore the associated color degrees of freedom can
completely thermalize.}
\vskip5pt
\item{$\underline{{\bf L} {\bf L} {\bf e}}$~ This monomial represents
a subspace of complex dimension three~\cite{gkm}. $D$-flatness
requires that the two ${\bf L}$s are from different generations, while
$F$-flatness requires that ${\bf e}$ belongs to the third generations
(therefore all the three lepton generations will be involved).
Eq.~(\ref{flatcoupling}) then implies that $\lambda_2 \simeq 10^{-2}$,
which kinematically blocks two-body inflaton decay via the ${\bf \Phi}
{\bf H}_u {\bf H}_d$ term. The decay can nevertheless proceed via the
${\bf \Phi} {\bf H}_u {\bf \Psi}$ term. However, since ${\Psi}$ cannot
develop a large VEV, we should actually rotate to the $\left\{{\bf
\Psi},{\bf L^{\prime}}_1,{\bf L^{\prime}}_2\right\}$ basis instead and
consider the ${\bf L^{\prime}}_1 {\bf L^{\prime}}_2 {\bf e}$
monomial. Such a change of basis has no impact on $D$-flatness but
will affect the $F$-flatness. The reason is that in general
$\lambda_e$ is not diagonal in the $\left\{{\bf \Psi},{\bf
L^{\prime}}_1,{\bf L^{\prime}}_2\right\}$ basis, and hence
$L^{\prime}_1$ and $L^{\prime}_2$ will be coupled to all flavors of
$e$. This can be circumvented if the inflaton dominantly couples to
one of the ${\bf L}_i$ and ${\bf \Psi}$ is mainly ${\bf L}_i$. More
specifically, a feasible curvaton candidate will be obtained along the
${\bf L}_2 {\bf L}_3 {\bf e}_1$ direction if ${\bf \Psi} \approx {\bf
L}_1$. For this flat direction we have $\lambda_1 = 0$ and $\lambda_3
\sim 10^{-5}$ (see Eq.~(\ref{flatcoupling})). This implies that
two-body inflaton decay will proceed via the ${\bf \Phi} {\bf H}_u
{\bf L}_1$ term without trouble. We note that this flat direction
completely breaks the electroweak symmetry $SU(2)_{\rm W} \times
U(1)_{\rm Y}$, while not affecting $SU(3)_{\rm C}$.  Therefore color
degrees of freedom can fully thermalize. In passing we note that this
monomial could be useful to generate primordial magnetic
field,~see~\cite{Magnetic}.}
\vskip5pt
\item{$\underline{{\bf L} {\bf L} {\bf d} {\bf d} {\bf d}}$: This
monomial represents a subspace of complex dimension
three~\cite{gkm}. $D$-flatness requires that the two ${\bf L}$s and
the three ${\bf d}$s are all from different generations. This implies
that that $\lambda_2 \simeq 10^{-2}$ and $\lambda_1 = \lambda_3 = 0$
(see Eq.~(\ref{flatcoupling})).  Therefore two-body inflaton decay via
the ${\bf \Phi} {\bf H}_u {\bf H}_d$ term will be kinematically
forbidden, but the decay can proceed via the ${\bf \Phi} {\bf H}_u
{\bf \Psi}$ term. Again note that we should rotate to the $\left\{{\bf
\Psi},{\bf L^{\prime}}_1,{\bf L^{\prime}}_2\right\}$ basis, and hence
the ${\bf L^{\prime}}_1 {\bf L^{\prime}}_2 {\bf d} {\bf d} {\bf d}$
direction will be the relevant direction. However this direction
breaks all of the SM gauge group. This results in late thermalization
of the Universe~\cite{Averdi2} and the absence of thermal effects
which, as a consequence, does not yield early oscillations of the flat
direction. This is an undesirable feature for our scenario in which,
the curvaton oscillations are triggered by thermal effects, as we will
discuss in the next Section. }
\vskip5pt
\end{itemize}

To summarize, after taking all considerations into account, we find
that the most promising candidates for the curvaton within MSSM are
the ${\bf u}_1 {\bf d} {\bf d}$ and ${\bf Q}_1 {\bf L^{\prime}} {\bf
d}$ (and possibly ${\bf L}_2 {\bf L}_3 {\bf e}_1$) flat directions.


\section{Importance of the $A$-term}
\label{CPA}

Let us now discuss the role of the $A$-term in constructing a curvaton
model. In models with gravity and anomaly mediation, SUSY breaking
results in the usual soft term, $m^2_0 {\vert \varphi \vert}^2$, in
the scalar potential with $m_0 \simeq 100~{\rm GeV}-1$ TeV. There is
also a new contribution arising from integrating out heavy modes
beyond the scale $M$, which usually induces non-renormalizable
superpotential terms of the form
\begin{equation}
W\sim \lambda_{n}\frac{\widetilde{\varphi}^n}{n M^{n-3}}\,,
\end{equation}
where $\widetilde\varphi$ denotes the superfield which comprises the
flat direction $\varphi$. In general $M$ could be the string scale,
below which we can trust the effective field theory, or $M = M_{\rm
P}$ (in the case of supergravity).

In addition, there are also inflaton-induced supergravity corrections
to the flat direction potential. All these terms provide a general
contribution to the flat direction potential which are of the
form~\cite{drt}
\begin{equation}
\label{mflat}
V(\varphi)\sim H^2M_{\rm P}^2 f\left(\frac{\varphi}{M_{\rm P}}\right)\,,
~~V(\varphi)\sim HM_{P}^3f\left(\frac{\varphi^{n}}{M_{P}^{n-3}}\right)\,,
\end{equation}
where $f$ is some function.  The first contribution usually gives rise
to a Hubble induced correction to the mass of the flat direction, $c_H
H^2|\varphi|^2$, with an unknown coefficient, $c_H$, which depends on
the nature of the K\"ahler potential. The second contribution is the
Hubble-induced $A$-term.

Note that $c_H$ can have either sign. If $c_H \sim 1$, the flat
direction mass is $> H$. It therefore settles at the origin during
inflation and remains there. Since $\vert \varphi \vert = 0$ at all
times, the flat direction will have no interesting consequences in
this case. The positive Hubble induced mass to the flat direction has
a common origin to the Hubble induced mass correction to the inflaton
in supergravity models. This is the well known
$\eta$-problem~\cite{eta}, which arises because of the canonical form
of the inflaton part of the K\"ahler potential. A large $\eta$
generically spoils slow roll inflation. In order to have a successful
slow roll inflation, one needs $\eta \ll 1$

The Hubble induced terms can be eliminated completely from inflation
and MSSM flat direction sectors completely if there is a shift
symmetry in the K\"ahler potentials\footnote{A nice realization of
chaotic inflation within supergravity is obtained by implementing a
shift symmetry~\cite{kyy}. If the inflaton K\"ahler potential has the
form $K = \left(\phi + \phi^{\ast}\right)^2/M^2_{\rm P}$, instead of
the minimal one $K = \phi^{\ast} \phi/M^2_{\rm P}$, the scalar
potential along the imaginary part of $\phi$ remains flat even for
Transplanckian field values. Therefore it can play the role of the
inflaton in a chaotic model. Note that a shift symmetry also ensures
that the (positive) Hubble induced corrections to the mass of flat
directions vanishes at the tree-level and the terms in
Eq.~(\ref{mflat}) disappear. It is also possible to realize chaotic
inflation for sub Planckian field values in supergravity. For example,
see the multi-axions~\cite{Shamit} driven assisted
inflation~\cite{Assist}.}, or, the Heisenberg
symmetry~\cite{Gaillard}.  Note that string theory generically gives
rise to No-scale type K\"ahler potential~\cite{NO-Scale} which
preserves Heisenberg symmetry at tree level.  In either cases the
inflaton and the MSSM sectors are free from Hubble-induced mass and
A-terms.  In the following we will consider examples where there are
{\it no Hubble-induced terms} in the potential.

The relevant part of the scalar potential is then given
by
\beq \label{nonren}
V = m^2_0 {\vert \varphi \vert}^2 +
\lambda^2_{n} \frac{\vert \varphi \vert^{2(n-1)}} {M_P^{2(n-3)}}+
\left(A\lambda_{n}\frac{\varphi^{n}}{M_P^{n-3}}+h.c.\right)\,,
\eeq
where $\lambda_n\sim {\cal O}(1)$ and $n \geq 4$. Note
that the low energy $A$-term is a dimensionful quantity, i.e., $A \sim
m_0\sim {\cal O}(100~{\rm GeV}-1~{\rm TeV})$, and depends on a
phase. As shown in~\cite{gkm}, all of the MSSM flat directions are
lifted by higher-order terms with $n \leq 9$. If a flat direction is
lifted at the superpotential level $n$, the VEV that it acquires
during inflation will also depend on the presence of the $A$-term.

However the angular direction of the potential being flat, during
inflation it obtains random fluctuations. There will be equally
populated domains of Hubble patch size where the phase of the
$A$-term is positive and negative. In either case during inflation,
the flat direction VEV is given by:
\begin{equation}
\label{VEV}
\varphi_{inf} \sim \left(m_0 M_P^{n-3}\right)^{1/n-2}\,.
\end{equation}
However there is a distinction between a positive and a negative phase
of the $A$-terms. The difference in dynamics arises after the end of
inflation. In the case of positive $A$-term the flat direction starts
rolling immediately, but in the case of a negative $A$-term, the flat
direction remains in a false vacuum for which the VEV is given by
Eq.~(\ref{VEV}). The mass of the flat direction around this false
minimum is very small compared to the Hubble expansion rate during
inflation, i.e. $(3n^2-9n+8)m_{0}^2\ll H_{inf}^2$ where $n>3$.  During
inflation the flat direction obtains quantum fluctuations whose
amplitude is given by Eq.~({\ref{cobe}).

The flat direction can exit such a metastable minimum only if thermal
corrections are taken into account. From now onwards we will assume
that the flat direction is locked in a false vacuum with a VEV
$\varphi_{inf}$ during and right after inflation.  In the next Section
we will discuss the various thermal corrections a flat direction obtains.


\section{Thermal corrections to the flat direction}

\label{TCF}

If part of the SM gauge group remains unbroken by the flat direction
VEV, the associated gauge fields and gauginos will not receive an
induced mass.  Together with the light particles and sparticles, they
reach full thermal equilibrium through gauge interactions (for related
studies, see refs.~\cite{therm1,therm2}). The most important processes
are $2 \rightarrow 2$ and $2 \rightarrow 3$ scatterings with gauge
boson exchange in the $t$-channel~\cite{therm1}. For massless gauge
bosons (as happens for the unbroken subgroup) these scatterings are
extremely efficient and lead to an almost instant thermalization of
particles upon their production in inflaton
decay~\cite{therm2}~\footnote{However particles carrying quantum
numbers associated with the broken gauge subgroup reach full
equilibrium much later. The reason is that corresponding gauge fields
(and gauginos) acquire a mass from the flat direction VEV which
suppresses the thermalization rate of these
particles~\cite{Averdi1,Averdi2}.}.

The presence of a thermal bath with a temperature $T$ affects the flat
direction dynamics. This happens through the back-reaction of fields
which are coupled to the flat direction~\cite{Thermal1,Thermal2}. The
flat direction VEV naturally induces a mass $y \varphi_{inf}$ to the
field which are coupled to it, where $y$ is a gauge or Yukawa
coupling.  Depending on whether $y\varphi_{inf} \leq T$ or $y
\varphi_{inf} > T$, different situations arise.

\vskip5pt
\noindent
\begin{itemize}
\item{$y \varphi_{inf} \leq T$: Fields which have a
 mass smaller than temperature are kinematically accessible to the
thermal bath. They will reach full equilibrium and result in a thermal
correction $V_{\rm th}$ to the flat direction potential
\beq \label{effmass1}
V_{\rm th} \sim + y^2 T^2 {\vert \varphi \vert}^2 .
\eeq
The flat direction then starts oscillating, provided that $y T >
H$~\cite{Thermal1}. For $H \leq \Gamma_{\rm d}$ the Universe is in a
radiation-dominated phase, and hence $T \propto H^{1/2}$. For $H >
\Gamma_{\rm d}$ the inflaton has not completely decayed and the plasma
from partial inflaton decay has a temperature $T \sim \left(H
\Gamma_{\rm d} M_{\rm P}^2\right)^{1/4}$~\cite{kt}, which implies that
$T \propto H^{1/4}$.  Therefore in both cases $y T > H$ will remain
valid once oscillations start.  }

\vskip 5pt
\noindent
\item{$y \varphi_{inf} > T$: Fields which have a mass larger
than temperature will not be in equilibrium with the thermal bath. For
this reason they are also decoupled from the running of gauge
couplings (at finite temperature). This shows up as a correction to
the free energy of gauge fields, which is equivalent to a logarithmic
correction to the flat direction potential~\cite{Thermal2}
\beq \label{effmass2}
V_{\rm th} \sim  \pm \alpha ~ T^4 {\rm ln} \left({\vert \varphi \vert}^2
\right),
\eeq
\setcounter{footnote}{0}
where $\alpha$ is a gauge fine structure constant. Decoupling of gauge
fields (and gauginos) results in a positive correction, while
decoupling of matter fields (and their superpartners) results in a
negative sign. The overall sign then depends on the relative
contribution of decoupled fields~\footnote{For example, consider the
${\bf H}_u {\bf H}_d$ flat direction. This direction induces large
masses for the top (s)quarks which decouples them from the thermal
bath, while not affecting gluons and gluinos.  Therefore this leads to
a positive contribution from the free energy of the gluons.}.
Obviously only corrections with a positive sign can lead to flat
direction oscillations around the origin (as we require). Oscillations
begin when the second derivative of the potential exceeds the Hubble
rate-squared which, from Eq.~(\ref{effmass2}), reads $\left(\alpha
T^2/\varphi_{inf}\right) > H$.}
\end{itemize}

Note that thermal effects of the first type require that fields which
have Yukawa couplings to the flat direction are in full equilibrium,
while those of the second type require the gauge fields (and gauginos)
be in full equilibrium. If all of the SM group is broken by flat
direction(s), the reheated plasma will reach full equilibrium much later
after the inflaton decay~\cite{Averdi2}. Instead it enters a long phase of
quasi-adiabatic evolution during which the plasma remains dilute. This
implies that fields which are coupled to the flat direction, and hence
have a large mass, decay quickly and are not produced again as the inverse
decays are inefficient. As a consequence, thermal effects, which crucially
depend on the presence of these degrees of freedom, will be weakened. This
is the reason why we want a subgroup of the SM gauge symmetry to
remain unbroken.

Let us now examine the thermal effects for the curvaton candidates
identified in Section III. Note that the instantaneous temperature of
the thermal bath from (partial) inflaton decay is $T \leq m_{\phi}$
for a perturbative inflaton decay. Therefore the kinematical condition
for inflaton decay via the ${\bf \Phi} {\bf H}_u {\bf H}_d$ and ${\bf
\Phi} {\bf H}_u {\bf \Psi}$ terms, given in Eqs.~(\ref{yukawabound1},
\ref{yukawabound2}) respectively, implies that only fields with a
Yukawa coupling $\lambda \leq 10^{-5}$ to the flat direction can be in
equilibrium with the thermal bath. We only have two possibilities:

\begin{itemize}
\item
$\underline {{\bf u}_1 {\bf d} {\bf d}}$: $SU(2)_{\rm W}$ remains
unbroken in this case. This implies that the cor\-res\-pon\-ding gauge
fields and gauginos, $H_u$ and $H_d$ (plus the Higgsinos) and the LH
(s)leptons reach full thermal equilibrium. The back-reaction of $H_u$
results in a thermal correction $\lambda^2_1 T^2 {\vert \varphi
\vert}^2 $, see Eq.~(\ref{flatcoupling}), with $\lambda_1 \sim
10^{-5}$. The free energy of the $SU(2)_{\rm W}$ gauge fields result
in a thermal correction $\sim + \alpha_{\rm W} T^4 {\rm ln}
\left({\vert \varphi \vert}^2 \right)$. Note that the sign is positive
since the flat direction induces a mass which is $> T$ (through the
$d$) for the LH (s)quarks but not the $SU(2)_{\rm W}$ gauge fields and
gauginos.

\item
$\underline {{\bf Q}_1 {\bf L}^{\prime} {\bf d}}$: An $SU(2)$ subgroup
of $SU(3)_{\rm C}$ is unbroken in this case. Therefore only the
corresponding gauge fields and gauginos plus some of the (s)quark
fields reach full thermal equilibrium. Then the back-reaction of ${\bf
u}_1$ and ${\bf d}_1$ results in a thermal correction $\left(
\lambda^2_1 + \lambda^2_2\right) T^2 {\vert \varphi \vert}^2$
according to Eq.(\ref{flatcoupling}), where $\lambda_1 \sim \lambda_2
\sim 10^{-5}$.  Note that logarithmic thermal corrections will not be
useful in this case as decoupling of a number of gluons (and gluinos)
from the running of strong gauge coupling results in a negative
contribution to the free energy of the unbroken part of $SU(3)_{\rm
C}$.
\end{itemize}

Therefore, within MSSM and from the point of view of thermal effects,
the ${\bf u}_1 {\bf d} {\bf d}$ flat direction is the most suitable
curvaton candidate\footnote{As pointed earlier, the ${\bf L}_2 {\bf
L}_3 {\bf e}_1$ flat direction can obtain a large VEV in the specific
case where ${\bf \Psi} = {\bf L}_1$. The $SU(3)_{\rm C}$ part of the
SM gauge symmetry remains unbroken for this flat direction, and hence
gluons, gluinos and (s)quarks will reach full equilibrium. We note
that neither of ${\bf L}$ and ${\bf e}$ are coupled to the color
degrees of freedom. This implies that there will be no $T^2 \varphi^2$
or logarithmic correction to the flat direction potential in this
case. This excludes the ${\bf L}_2 {\bf L}_3 {\bf e}_1$ flat direction
from being a successful curvaton candidate.}


\section{${\bf u_1dd}$ as the MSSM curvaton}
\label{udd}

The flat directions represented by the ${\bf udd}$ monomial are partly
lifted by $n=4$ superpotential terms, and eventually lifted at the
$n=6$ level~\cite{gkm}. The former are lifted by the ${\bf u d d e}$
superpotential term which implies a zero $A$-term because ${\bf udd}$
achieves a large VEV which renders a large mass for ${\bf e}$ and
consequently $\langle {\bf e}\rangle =0$. Therefore the scalar
potential along these directions will have no metastable minimum. The
latter, on the other hand, are lifted by the $\left({\bf u d
d}\right)^2$ superpotential term which also induces a non-zero
$A$-term (as required). As we shall see, for these directions
sufficient thermal corrections are induced to lift the flat direction
from its false minimum.

The only channel for two-body inflaton decay is through the ${\bf \Phi
H_u\Psi}$ term. The existence of two generations of ${\bf d}$ in the
flat direction kinematically blocks two-body decay via the other
renormalizable term ${\bf \Phi H_uH_d}$ (see the discussion in
Section~\ref{Curvatonc}). Note that the $SU(2)_{\rm W}$ gauge fields
and gauginos are massless as this subgroup remains unbroken by the
${\bf u_1 d d}$ direction. Therefore degrees of freedom which carry
$SU(2)_{\rm W}$ gauge quantum numbers, and their induced mass by the
curvaton is $\ls T$, fully thermalize instantly (compared with the
Hubble expansion rate) via $2 \rightarrow 2$ and $2 \rightarrow 3$
scatterings with $SU(2)_{\rm W}$ gauge bosons in the
t-channel~\cite{therm2,Averdi2}. On the other hand, the $SU(3)_{\rm C}
\times U(1)_{\rm Y}$ gauge fields and gauginos, plus the right-handed
(s)quarks and (s)leptons, reach full equilibrium much later as their
gauge interactions are suppressed by the flat direction induced
mass~\cite{Averdi2}. In the following $T$ refers to the temperature of
$SU(2)_{\rm W}$ degrees of freedom which are in {\it full} thermal
equilibrium.

From the discussion in the previous Section it follows that thermal effects
result in a correction to the effective thermal flat direction potential
given by
\beq
\label{thermeff}
V_{\rm th} \sim \lambda^2_1 T^2 {\vert \varphi \vert}^2 +
\alpha_{\rm W} T^4 {\rm ln} \left({\vert \varphi \vert}^2\right),
\eeq
where $\lambda_1 \sim 10^{-5}$ and $\alpha_{\rm W} \sim 10^{-2}$.

Note that the first and foremost condition for lifting the flat
direction from its metastable minimum is that $V_{\rm th} > m^2_0
\varphi^2_{inf}$. On the other hand, a perturbative inflaton decay
yields a radiation-dominated Universe whose temperature is $T \leq
T_{\rm R} \leq m_{\phi}$. The reheat temperature $T_{\rm R} \sim
\sqrt{\left(\Gamma_{\rm d} M_{\rm P}\right)}$ is the largest
temperature of the Universe in a quasi-thermal radiation-dominated
phase, where $\Gamma_{\rm d}$ is the inflaton decay rate given in
Eq.~(\ref{Gammad}).

If $T_{\rm R} \sim m_{\phi}$, the first term on the right-hand side of
Eq.~(\ref{thermeff}) dominates over the flat direction energy density
given by Eq.~(\ref{nonren}) for $n=6$ right after the inflaton has
completely decayed. Note that $\varphi_{inf} \sim 10^5 m_{\phi}$ from
Eq.~(\ref{cobe}). Note however that very soon we will have $T <
\lambda_1 \varphi_{inf}$ since the Hubble expansion rate is gradually
decreasing and the thermal contribution, i.e. $\lambda^2_1 T^2 {\vert
\varphi \vert}^2$, becomes ineffective, since the fields with coupling
$\lambda_1$ to the ${\bf u}_1 {\bf d} {\bf d}$ direction have a mass
$> T$ and hence drop out of quasi-thermal equilibrium. Eventually the
logarithmic thermal correction would take over very quickly, even
if~\footnote{Obviously the $T^2 {\vert \varphi \vert}^2$ correction
will not arise at all if $T_{\rm R} \ll m_{\phi}$, since in this case
we will have $\lambda_1 \varphi_{inf} \gg T$.} $T_{\rm R} \approx
m_{\phi}$. For this reason we focus on the second term of
Eq.~(\ref{thermeff}) in the discussion below.

First of all we need $V_{\rm th} > m^2_0 \varphi^2_{inf}$, so that
the thermal effects will overcome the potential barrier. This leads
to the requirement
\beq \label{cond1}
\alpha_{\rm W} T^4 > m^2_0 \varphi^2_{inf}.
\eeq
Second, the thermal mass should trigger flat direction oscillations. If
oscillations do not begin, the flat direction simply sits at a field
value $\simeq \varphi_{inf}$ while $V_{\rm th}$ is redshifted
$\propto H^2$ in a radiation-dominated Universe. The flat direction
will be trapped again in the metastable minimum when $V_{\rm th} <
m^2_0 \varphi^2_{inf}$.

In order for the flat direction oscillations to start we must have
$d^2V_{\rm th}/d\vert \varphi \vert^2 > H^2$. This leads
to the condition
\beq \label{cond2}
\alpha^{1/2}_{\rm W} \frac{T^2} {\varphi_{inf}} >H(T)\,,
\eeq
which always holds in a radiation-dominated phase where $H \simeq
T^2/M_{\rm P}$ (note that $\varphi_{inf} \ll M_{\rm P}$). This implies
that the ${\bf u}_1 {\bf d} {\bf d}$ direction starts oscillating once
the condition given in Eq.~(\ref{cond1}) is satisfied. This happens,
 when temperature of the Universe is given by
\beq
\label{tosc}
T_{osc} \sim \left(\frac{\varphi_{inf}} {10^{14}~{\rm GeV}}\right)^{1/2} \times
10^9~{\rm GeV}.
\eeq
After taking into account of the constraints $T_{osc} \leq m_{\phi}$,
which holds for a perturbative inflaton decay, and $m_{\phi} \sim
10^{-5} \varphi_{inf}$, which arises from the amplitude of the CMB
anisotropy, we obtain
\beq \label{ampcond}
\varphi_{inf} \geq 10^{14} ~ {\rm GeV}.
\eeq
Interestingly enough for directions which are lifted at the $n=6$
superpotential level we have $\varphi_{inf} \sim 3 \times 10^{14}$
GeV, see Eq.~(\ref{VEV}).  Then the inflaton mass is determined from
Eq.~(\ref{cobe}) to be $m_{\phi} \sim 3 \times 10^9$ GeV. Also the
requirement that $T_{\rm R} \geq T_{osc}$ (a radiation-dominated
Universe is formed only after the inflaton has completely decayed)
restricts the inflaton decay rate according to
Eqs.~(\ref{Gammad}),(\ref{tosc}). A lower bound on the renormalizable
inflaton coupling is $h \geq 10^{-4}$ from Eqs.~({\ref{inf}),
(\ref{Gammad}), which is perfect for the validity of a perturbative
inflaton decay~\cite{Averdi2}. Note that the flat direction VEV
prevents non-perturbative inflaton decay as described in
Ref.~\cite{AM}.

Last but an important reminder to the readers is that the evaporation
of the flat direction is not an important issue in our case. The
interaction term that leads to evaporation is of the form $\lambda_1^2
\varphi^2 \chi^2$, with $\varphi$ being the flat direction and, $\chi$
collectively denotes the fields in thermal equilibrium. The rate for
evaporation is then given by:
\beq
\Gamma \simeq \lambda_1^4 \frac{n_{\chi}}{E_{\chi} E_{\varphi}}\,.
\eeq
where $n_{\chi} \sim T^3$, $E_\chi \sim T$ and $E_\varphi \sim
T^2/\langle\varphi\rangle$. Note that the latter being the mass of flat
direction due to logarithmic thermal corrections, where $\langle
\varphi\rangle$ is the the amplitude of the flat direction
oscillations. We then find:
\beq \Gamma \sim \lambda_1^4 \langle\varphi\rangle << H(T)\,.  \eeq
since $\lambda_1\sim 10^{-5}$ for $u_1 d d$ flat direction. It takes a
large number of oscillations for the flat directions to evaporate.
Further note that during radiation domination, the flat direction VEV
scales like: $\langle \varphi\rangle \propto H^{3/4}$, and $\langle
\varphi\rangle \propto H$ during matter domination and/or when the
flat flat direction oscillates. Therefore, thermal evaporation is not
an important threat before the curvaton oscillations dominate the
Universe~\footnote{Along the same lines, it is also true that in the
case of a quadratic thermal correction, the evaporation rate is again
subdominant compared to the Hubble expansion rate.}.

Hence we conclude that the ${\bf udd}$ flat direction is a realistic
curvaton candidate in the MSSM. At the time of its decay, it dominates
the energy density of the universe so that its isocurvature
perturbations are converted into the observed curvature
perturbations in CMB. During the last stages
of the ${\bf udd}$ oscillations, the flat direction generically
fragments into $Q$-balls~\cite{Q-balls}. The $Q$-balls eventually
decay into LSPs through surface evaporation, which can lead to the
observed dark matter. Since the $Q$-balls behave like non-relativistic
matter, they again start dominating the radiation energy density until
they completely evaporates. However, this does not modify the
super-Hubble perturbations.  The fragmentation of the flat direction
into $Q$-balls is strictly a sub-Hubble process.


\section{Conclusion}
\label{lopetus}

We have considered thermal corrections in the presence of a MSSM flat
direction condensate, which gives rise to gauge symmetry breaking and
a slowing down of thermalization rates. We have paid particular
attention to the coupling of the inflaton to ordinary SM matter and
identified among the nearly $300$ MSSM flat directions the ${\bf udd}$
direction lifted by $n=6$ non-renormalizable terms as a successful
MSSM curvaton candidate that survives the constraints of energy
density dominance and the condensate non-dissociation by the ambient
plasma.  To our knowledge this is the first paper where thermal
corrections to the MSSM flat direction curvaton are accounted for
properly.

The ${\bf u_1dd}$ direction provides masses for the gauge bosons of
$SU(3)_{\rm C} \times U(1)_{\rm Y}$ while leaving $SU(2)_{\rm W}$
unbroken.  It is thus only the $SU(2)_{\rm W}$ degrees of freedom that
contribute to the thermal correction of the flat direction potential,
as given in Eq. (\ref{thermeff}).  The dominance of ${\bf udd}$ over
thermal plasma comes about because of two important factors. First,
the condensate has an initial amplitude which is large,
i.e. $\varphi_{inf}\sim 3\times 10^{14}$~GeV.  Second, there is an
$A$-term which creates a false vacuum during and after inflation which
traps the flat direction VEV for a sufficiently long time.

The CMB fluctuations imparted by ${\bf udd}$ not only have
cosmological implications but also astrophysical ones. The flat
direction is also a well motivated candidate for generating the cold
dark matter through $Q$-ball evaporation. Future collider-based and
astrophysical experiments will hopefully pin down the nature of dark
matter and the physics beyond the SM, but for the time being we are
assured that minimal supersymmetric Standard Model can provide a
cosmologically viable curvaton candidate, ${\bf udd}$ responsible for
the CMB fluctuations and reheating, and possibly also accounting for
the dark matter.

As in all curvaton scenarios, the spectral index $n_s$ for the ${\bf
udd}$ curvaton is very close to 1 whereas WMAP 3-year data indicates
\cite{WMAP3yr} for the CMB power spectrum
$n_s=0.951^{+0.015}_{-0.019}$.  However, in case of the ${\bf udd}$
curvaton the spectral index of the power spectrum depends also on the
yet unspecified inflaton sector with possible tensor perturbations.


\ack

The work of R.A. is supported by the Natural Sciences and Engineering
Research Council of Canada (NSERC). K.E. is supported in part by the
Academy of Finland grant no. 205800. A.M. would like to thank CERN,
University of Padova, Perimeter Institute and McGill University for
their kind hospitality where parts of this project were carried out.



\section*{References}

\begin{thebibliography}{31}
%

\bibitem{Enqvist-REV}
 K.~Enqvist and A.~Mazumdar,
  Phys.\ Rept.\  {\bf 380}, 99 (2003)
  [arXiv:hep-ph/0209244].
M. Dine and A. Kusenko, Rev. Mod. Phys. {\bf 76}, 1 (2004)
[arXiv:hep-ph/0303065].

\bibitem{Averdi1}
  R.~Allahverdi and A.~Mazumdar,
  arXiv:hep-ph/0505050.

\bibitem{Averdi2}
R.~Allahverdi and A.~Mazumdar,
  arXiv:hep-ph/0512227.

\bibitem{Enqvist1}
K.~Enqvist, S.~Kasuya and A.~Mazumdar,
  Phys.\ Rev.\ Lett.\  {\bf 90}, 091302 (2003)
  [arXiv:hep-ph/0211147].


\bibitem{Enqvist2}
 K.~Enqvist, A.~Jokinen, S.~Kasuya and A.~Mazumdar,
  Phys.\ Rev.\ D {\bf 68}, 103507 (2003)
  [arXiv:hep-ph/0303165].


\bibitem{Enqvist3}
 K.~Enqvist, S.~Kasuya and A.~Mazumdar,
  Phys.\ Rev.\ Lett.\  {\bf 93}, 061301 (2004)
  [arXiv:hep-ph/0311224].
 K.~Enqvist, A.~Mazumdar and A.~Perez-Lorenzana,
  Phys.\ Rev.\ D {\bf 70}, 103508 (2004)
  [arXiv:hep-th/0403044].
 K.~Enqvist, A.~Mazumdar and M.~Postma,
  Phys.\ Rev.\ D {\bf 67}, 121303 (2003)
  [arXiv:astro-ph/0304187].


\bibitem{Postma}
 M.~Postma,
  Phys.\ Rev.\ D {\bf 67}, 063518 (2003)
  [arXiv:hep-ph/0212005].
S.~Kasuya, M.~Kawasaki and F.~Takahashi,
  Phys.\ Lett.\ B {\bf 578}, 259 (2004)
  [arXiv:hep-ph/0305134].
A. Mazumdar and M. Postma,
Phys. Lett. B {\bf 573}, 5 (2003) [Erratum-ibid. B {\bf 585}, 295 (2004)]
[arXiv:astro-ph/0306509].
 M.~Postma and A.~Mazumdar,
  JCAP {\bf 0401} (2004) 005
  [arXiv:hep-ph/0304246].
R. Allahverdi, Phys. Rev. D {\bf 70}, 043507 (2004) [arXiv:astro-ph/0403351].
M.~Ikegami and T.~Moroi,
  Phys.\ Rev.\ D {\bf 70}, 083515 (2004)
  [arXiv:hep-ph/0404253].
A.~Mazumdar,
  Phys.\ Rev.\ Lett.\  {\bf 92}, 241301 (2004)
  [arXiv:hep-ph/0306026].
A.~Mazumdar and A.~Perez-Lorenzana,
  Phys.\ Rev.\ Lett.\  {\bf 92}, 251301 (2004)
  [arXiv:hep-ph/0311106].
 A.~Mazumdar and A.~Perez-Lorenzana,
  Phys.\ Rev.\ D {\bf 70}, 083526 (2004)
  [arXiv:hep-ph/0406154].



\bibitem{Sloth}
K.~Enqvist and M.~S.~Sloth,
  Nucl.\ Phys.\ B {\bf 626}, 395 (2002)
  [arXiv:hep-ph/0109214].
A.~D.~Linde and V.~F.~Mukhanov,
  Phys.\ Rev.\ D {\bf 56}, 535 (1997)
  [arXiv:astro-ph/9610219].

\bibitem{Lyth-Wands}
D.~H.~Lyth and D.~Wands,
  Phys.\ Lett.\ B {\bf 524}, 5 (2002)
  [arXiv:hep-ph/0110002].
 D.~H.~Lyth, C.~Ungarelli and D.~Wands,
  Phys.\ Rev.\ D {\bf 67}, 023503 (2003)
  [arXiv:astro-ph/0208055].
D.~H.~Lyth,
  arXiv:astro-ph/0602285.

\bibitem{Moroi}
 T.~Moroi and T.~Takahashi,
  Phys.\ Lett.\ B {\bf 522}, 215 (2001)
  [Erratum-ibid.\ B {\bf 539}, 303 (2002)]
  [arXiv:hep-ph/0110096].

\bibitem{Thermal1}
R. Allahverdi, B. A. Campbell and J. R. Ellis, Nucl. Phys. B {\bf 579},
355 (2000) [arXiv:hep-ph/0001122].


\bibitem{Thermal2}
A. Anisimov and M. Dine, Nucl. Phys. B {\bf 619}, 729 (2001)
[arXiv:hep-ph/0008058];
A. Anisimov, Phys. Atom. Nucl. {\bf 67}, 640 (2004) [arXiv:hep-ph/0111233].

\bibitem{EKM}
K.~Enqvist, S.~Kasuya and A.~Mazumdar,
  Phys.\ Rev.\ Lett.\  {\bf 89}, 091301 (2002)
  [arXiv:hep-ph/0204270].
K.~Enqvist, S.~Kasuya and A.~Mazumdar,
  Phys.\ Rev.\ D {\bf 66}, 043505 (2002)
  [arXiv:hep-ph/0206272].
R.~Allahverdi, R.~Brandenberger and A.~Mazumdar,
  Phys.\ Rev.\ D {\bf 70}, 083535 (2004)
  [arXiv:hep-ph/0407230].

\bibitem{AM} 
R. Allahverdi and A. Mazumdar, ``Towards a successful
reheating within supersymmetry'',  arXiv:hep-ph/0603244.

\bibitem{Asko-Assist}
A.~Jokinen and A.~Mazumdar,
  Phys.\ Lett.\ B {\bf 597}, 222 (2004)
  [arXiv:hep-th/0406074].


\bibitem{Asko-multi}
 K.~Enqvist, A.~Jokinen and A.~Mazumdar,
  JCAP {\bf 0401}, 008 (2004)
  [arXiv:hep-ph/0311336].

\bibitem{sninfl}
H. Murayama, H. Suzuki, T. Yanagida and J. Yokoyama, Phys. Rev. Lett.
{\bf 70}, 1912 (1993); H. Murayama, H. Suzuki, T. Yanagida and J. Yokoyama,
Phys. Rev. D {\bf 50}, 2356 (1994) [arXiv:hep-ph/9311326].




\bibitem{gkm}
T. Gherghetta, C. Kolda and S. P. Martin, Nucl. Phys. B {\bf 468},
37 (1996) [arXiv:hep-ph/9510370].


\bibitem{Magnetic}
 K.~Enqvist, A.~Jokinen and A.~Mazumdar,
  JCAP {\bf 0411}, 001 (2004)
  [arXiv:hep-ph/0404269].




\bibitem{drt}
M.~Dine, L.~Randall and S.~Thomas, Phys. Rev. Lett. {\bf 75}, 398 (1995)
[arXiv:hep-ph/9503303].
M. Dine, L. Randall and S. Thomas, Nucl. Phys. B {\bf 458}, 291 (1996)
[arXiv:hep-ph/9507453].


\bibitem{eta}
M. Dine, W. Fischler, and D. Nemeschansky, Phys. Lett. B {\bf 136}, 169 (1984);
G. D. Coughlan, R. Holman, P. Ramond, and G. G. Ross, Phys. Lett. B {\bf 140},
44 (1984); 
A. S. Goncharov, A. D. Linde, and M. I. Vysotsky, Phys. Lett. B
{\bf 147}, 279 (1984); 
O. Bertolami, and G. G. Ross, Phys. Lett. B {\bf 183},
163 (1987);
E. J. Copeland, A. R. Liddle, D. H. Lyth, E. D. Stewart, and D. Wands,
Phys. Rev. D {\bf 49}, 6410 (1994)
[arXiv:astro-ph/9401011].




\bibitem{kyy}
M.~Kawasaki, M.~Yamaguchi and T.~Yanagida,
Phys.\ Rev.\ Lett.\  {\bf 85}, 3572 (2000)
[arXiv:hep-ph/0004243].

\bibitem{Shamit}
S.~Dimopoulos, S.~Kachru, J.~McGreevy and J.~G.~Wacker,
arXiv:hep-th/0507205.





\bibitem{Assist}
 A.~R.~Liddle, A.~Mazumdar and F.~E.~Schunck,
  Phys.\ Rev.\ D {\bf 58}, 061301 (1998)
  [arXiv:astro-ph/9804177].
 E.~J.~Copeland, A.~Mazumdar and N.~J.~Nunes,
  Phys.\ Rev.\ D {\bf 60}, 083506 (1999)
  [arXiv:astro-ph/9904309].
 A.~Mazumdar, S.~Panda and A.~Perez-Lorenzana,
  Nucl.\ Phys.\ B {\bf 614}, 101 (2001)
  [arXiv:hep-ph/0107058].



\bibitem{Gaillard}
M.~K.~Gaillard, H.~Murayama and K.~A.~Olive, Phys. Lett. B {\bf 355}, 71
(1995) [arXiv:hep-ph/9504307].


\bibitem{NO-Scale}
 A.~B.~Lahanas and D.~V.~Nanopoulos,
  Phys.\ Rept.\  {\bf 145}, 1 (1987).

\bibitem{therm1}
J. McDonald, Phys. Rev. D {\bf 61}, 083513 (2000) [arXiv:hep-ph/9909467];
R.~Allahverdi,
Phys.\ Rev.\ D {\bf 62}, 063509 (2000) [arXiv:hep-ph/0004035].



\bibitem{therm2}
S. Davidson and S. Sarkar, JHEP {\bf 0011}; 012 (2000) [arXiv:hep-ph/0009078];
R. Allahverdi and M. Drees, Phys. Rev. D {\bf 66},
063513 (2002) [arXiv:hep-ph/0205246];
P. Jaikumar and A. Mazumdar, Nucl. Phys. B {\bf 683},
264 (2004) [arXiv:hep-ph/0212265].


\bibitem{kt}
E. W. Kolb and M. S. Turner, {\it The Early Universe}, Addison-Wesely,
New York 1990.


\bibitem{Q-balls}
A.~Kusenko,
  Phys.\ Lett.\ B {\bf 405}, 108 (1997)
  [arXiv:hep-ph/9704273].
  A.~Kusenko,
  Phys.\ Lett.\ B {\bf 404}, 285 (1997)
  [arXiv:hep-th/9704073].
 A.~Kusenko and M.~E.~Shaposhnikov,
  Phys.\ Lett.\ B {\bf 418}, 46 (1998)
  [arXiv:hep-ph/9709492].
 K.~Enqvist and J.~McDonald,
  Phys.\ Lett.\ B {\bf 425}, 309 (1998)
  [arXiv:hep-ph/9711514].
 K.~Enqvist and J.~McDonald,
  Nucl.\ Phys.\ B {\bf 538}, 321 (1999)
  [arXiv:hep-ph/9803380].

\bibitem{WMAP3yr}
D.N. Spergel, et.al., astro-ph/0603449.


\end{thebibliography}
\end{document}